\documentclass{article}
\usepackage{dafne99pro}
\begin{document}
\title{ 
LIGHT MESON SPECTROSCOPY: \\
RECENT DEVELOPMENTS and DAFNE
}
\author{
T. Barnes \\
{\em Physics Division, Oak Ridge National Laboratory} \\ 
{\em Oak Ridge, TN 37831-6373, USA } \\ 
{\em Department of Physics, University of Tennessee  } \\ 
{\em Knoxville, TN 37996-1501, USA} \\ 
{\em Institut f\"ur Theoretische Kernphysik der Universit\"at Bonn} \\ 
{\em Bonn D-53115, Germany} \\ 
{\em Institut f\"ur Kernphysik, Forschungszentrum J\"ulich} \\ 
{\em J\"ulich D-52425, Germany} \\ 
}
\maketitle
\baselineskip=11.6pt

\begin{abstract}
In this contribution I discuss
recent developments in light meson spectroscopy, and note specific
areas in which DAFNE is an especially appropriate tool for
future experiments. One topic of special relevance is the
spectroscopy of excited vector mesons;
quite narrow vector hybrids are predicted by the
flux-tube model, which could be produced by DAFNE when operating in
the $M_{e^+e^-}\approx 1.5-2$~GeV range. A second
topic, which would be appropriate for a later date because it requires 
a rather higher beam energy, is the production
of C=(+) mesons in $\gamma\gamma$ collisions. 

\end{abstract}
\baselineskip=14pt

\section{Introduction}

The last few years have seen rapid and exciting developments 
in light meson spectroscopy, largely as a result of the analysis 
of high-statistics 
experiments
using hadron beams. The most notable discoveries have come from 
studies of
$P\bar P$ annihilation
at LEAR
and 
$\pi^- P$
at the AGS (BNL) and VES (Serpukhov).
In both processes we have seen that detailed amplitude analyses of
high-statistics events samples (ca.1M events) have made possible
the identification of very interesting parent resonances 
in otherwise relatively mundane final states such as $3\pi$.
This has 
led for example to the discovery of a glueball candidate 
in $3\pi^o$ and an exotic
hybrid candidate in $(3\pi)^-$. 
Concurrently we have seen impressive progress in the
study of conventional $q\bar q$ mesons (which must be 
identified as a background to more unusual resonances), 
and at this meeting we have heard important 
new results from VEPP which appear to confirm the 
predictions of Close, Isgur and Kumano for a
$K\bar K$-molecule 
assignment for the scalars $f_0(980)$ and $a_0(980)$. 
In this case at least, progress has come from an $e^+e^-$ facility 
rather than a hadronic one.
In this introduction I will give a brief summary of the status of 
the various sectors of meson spectroscopy, and then discuss two areas
in which DAFNE can make very important 
contributions, excited vectors and C=(+) mesons. 

\section{Recent developments in light meson spectroscopy.}

\subsection{Glueballs}

\begin{figure}[t]
 \vspace{9.0cm}
\includegraphics{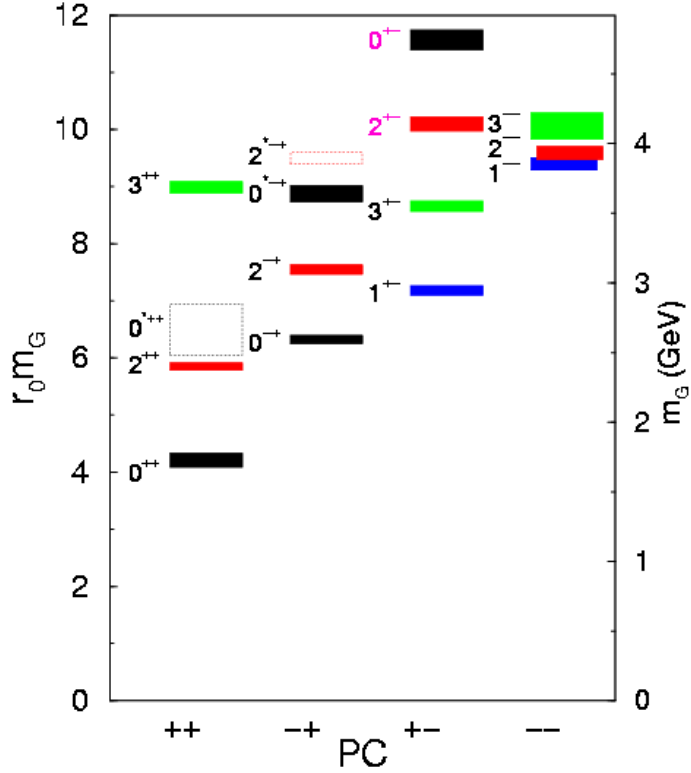}
 \caption{\it
The spectrum of glueballs found by Morningstar and Peardon
in pure glue LGT.\cite{Colin} 
The
lowest scalar has a predicted mass of 1.73(5)(8)GeV.
    \label{fig1} }
\end{figure}

\begin{figure}[t]
 \vspace{9.0cm}
\includegraphics{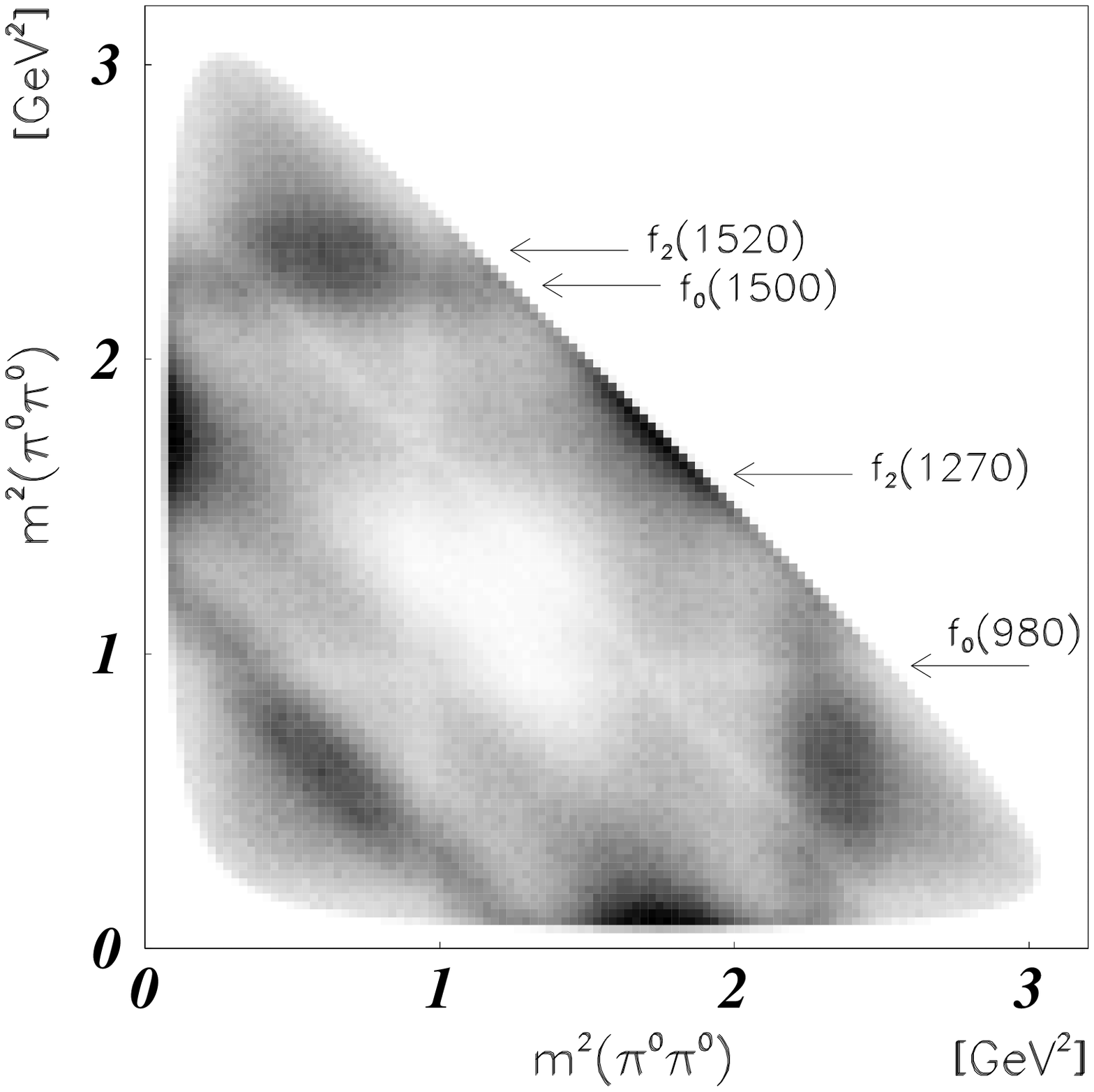}
 \caption{\it
The scalar glueball candidate $f_0(1500)$ observed
by the
Crystal Barrel Collaboration\cite{CBwebsite}
in $P\bar P \to \pi^o(2\pi^o$).
    \label{fig2} }
\end{figure}

The gluonic degree of freedom in QCD leads to more physical resonances 
than are predicted by the naive $q\bar q$ quark model.
Pure-glue ``glueball" states
have been studied using many 
theoretical approaches, 
the most recent and (presumably) the most 
accurate of which is lattice gauge
theory (LGT). 
In recent years LGT
has largely displaced other theoretical methods 
for treating these most unfamiliar of hadrons. 
A recent high-statistics LGT study of the glueball spectrum 
to ca.4~GeV has been reported
by
Morningstar and Peardon\cite{Colin}(see Fig.1); 
for other recent discussions of glueballs and LGT 
see Teper\cite{Teper_LGT}
and Michael.\cite{Michael_LGT}
The lattice predicts that the lightest (assumed unmixed with $q\bar q$)
glueball is a scalar, with a mass of about 1.7~GeV. Additional
glueballs lie well above 2 GeV, with a 
$0^{-+}$
and a
$2^{++}$
appearing at masses 
of $\approx 2.4-2.6$ GeV.  Spin-parity exotic glueballs
are expected at rather higher masses; in the 
Morningstar and Peardon study the lightest exotic glueball was 
found to be a
$2^{+-}$ 
at just above
4~GeV. 
For experimental 
studies of meson spectroscopy
below ca.2.2~GeV, the subject
of glueballs thus 
reduces to the search for an extra scalar. 

Scalars unfortunately comprise the most obscure part of the spectrum,
and there are at least three states that might {\it a priori} be identified
with a scalar glueball, the $f_0(1370)$, the LEAR state
$f_0(1500)$\cite{AC} and the
$\psi$ radiative candidate $f_0(1710)$.\cite{SVW} 

There are outstanding problems with each of these assignments.
In view of LGT mass predictions the $f_0(1500)$ and $f_0(1710)$
appear most plausible, but neither of these states shows the
flavor-blind pattern of decay couplings naively expected for
a flavor-singlet glueball. 
The $f_0(1500)$ as seen by Crystal Barrel in $\pi^o\pi^o$ is shown
in Fig.2.
The results of some analyses, taken from
the 1998 PDG, are shown in Table~1. Although essentially all these
numbers are controversial, it is clear that the $\pi\pi / K\bar K$
branching ratios of the $f_0(1500)$ and $f_0(1710)$ are both far
from the approximate equality expected for a flavor-singlet. 
We also note that the two lighter states have large $4\pi$ modes,
which have not been considered in glueball decay models.
\begin{table}
\centering
\caption{\it 
Some two-pseudoscalar
branching fractions 
of $f_0$ states quoted by the
PDG.\cite{PDG98}
}
\vskip 0.1 in
\begin{tabular}{|l|c|c|c|c|c|} 
\hline
Mode:
& $\pi\pi$  
& $K\bar K$  
&  $\eta\eta$  
& $\eta\eta'$  
&  $\eta'\eta'$  
\\
\hline
\hline
singlet/$k_f$:
& 3 
& 4
& 1
& 0
& 1
\\
\hline
\hline
$B_i$ (expt.):
&  
& 
& 
& 
& 
\\
\hline
$f_0(1370)$
&  $ 26\% \ (9\% )$
&  $ 35\% \ (13\% )$
&   seen
&   -
&   -
\\
\hline
$f_0(1500)$
&  $ 45.4\% \ (10.4\% )$
&  $ 4.4\%  \ (2.4\% )$ 
&  seen
&  seen
&  -
\\
\hline
$f_0(1710)$
&  $ 3.9\% {+0.2\% \atop -2.4\% }$
&  $ 38\% {+9\% \atop -19\% }$
&  $ 18\% {+3\% \atop -13\% }$
&  -
&  -
\\
\hline
\end{tabular}
\label{table0}
\end{table}

The $K\bar K$ mode of the $f_0(1500)$ is difficult to
isolate, but appears to be weaker than one would expect for flavor-singlet
couplings to $\pi\pi$, $K\bar K$ and $\eta\eta$. 
Conversely, the $f_0(1710)$ has a strong $K\bar K$ mode but a weak
$\pi\pi$ coupling. The determination of the 
$K\bar K$ branching fraction of the $f_0(1500)$ has 
recently been reanalysed by
Ableev {\it et al,}\cite{Ableev} who find a much larger branching
fraction than quoted in Table~1, but still rather smaller than 
expected for a flavor singlet.
Several models, for example that of 
Amsler and Close,\cite{AC} 
invoke important $n\bar n \leftrightarrow G
\leftrightarrow s\bar s$ mixing to explain the 
observed branching fractions
these scalar states, 
so the scalar glueball basis state 
may actually be distributed over
several physical resonances.
In the final section 
we will discuss 
how this
possibility could be tested at an $e^+e^-$ facility.

For completeness we note that BES has reported evidence
for a possible narrow state in several channels,
including $P\bar P$, 
$\pi\pi$, $K\bar K$ and $\eta\eta$, at about 2.2~GeV.\cite{BES}
Although one does expect a tensor glueball not far above this mass, and
the narrow glueball candidate $f_0(1500)$ 
suggests that the tensor glueball might have a narrow width,
the statistical significance of the reported signals 
near 2.2~GeV is rather low. 
Another problem is that
the Crystal Barrel has shown that the $P\bar P$ and $\eta\eta$ modes
cannot both be as large as claimed by BES,
since the state does not appear with the corresponding strength in
$P\bar P \to \eta\eta$. This state clearly
``needs confirmation".

\subsection{Hybrid Mesons}

In addition to glueballs, we also expect the glue degree of
freedom to lead to ``hybrid mesons" in which the $q\bar q$ pair
is combined with glue in an excited state.
Hybrids are especially attractive experimentally, because they span
flavor nonets (so they can be searched for in many flavor channels),
and have ``exotic" 
$J^{PC}$  combinations such as
$1^{-+}$ that are forbidden to $q\bar q$ states.
(Hybrids span {\it all} $J^{PC}$ quantum numbers, both exotic and non-exotic.)
The 
$J^{PC}$ content of the
lowest-lying hybrid multiplet 
is model dependent:
The lowest-lying exotics in this first hybrid multiplet 
according to the 
flux-tube model are
\begin{equation}
J^{PC}({\rm lightest\  flux}-{\rm tube\ hybrid\ exotics}) =
0^{+-}, 1^{-+}, 2^{+-}
\label{eq1} 
\end{equation}
and are expected to be approximately degenerate.
In contrast, in the bag model the lightest hybrid multiplet only has
the single exotic
\begin{equation}
J^{PC}({\rm lightest\  bag-model\ hybrid\ exotic})  =
1^{-+}
\ .
\label{eq2}
\end{equation}
The difference is due to assumptions about confinement; the bag model 
has a confining boundary condition that discriminates between color electric
and magnetic fields, which gives a TM $(1^-)$ gluon more energy than 
TE $(1^+)$. The flux-tube model in contrast simply has a spatially excited 
interquark string and makes no reference to color field vectors. 
(Preliminary
LGT results found the $1^{-+}$ hybrid at a significantly 
lower mass than the
$0^{+-}$,\cite{MILC} 
as expected in the bag model but not the flux-tube model;
more recent results by the same collaboration now find the
$1^{-+}$ and $0^{+-}$ exotic hybrids
rather closer in mass.\cite{MILC2})
The mass of the lightest hybrid meson multiplet is expected by theorists
to be near 1.9~GeV. The bag model typically finds a somewhat lower scale of
ca.1.5~GeV, which is now deprecated because it disagrees with LGT.
This 1.9~GeV estimate was originally due to the flux tube 
model,\cite{IKP,BCS} and has been (approximately) 
confirmed by recent LGT 
studies, which find a mass of about 2.0~GeV for the lightest
hybrid.\cite{hybr_latt} 
For a recent review of LGT predictions
for these states see Michael.\cite{Michael_LGT}

\begin{figure}[t]
 \vspace{6.0cm}
\includegraphics{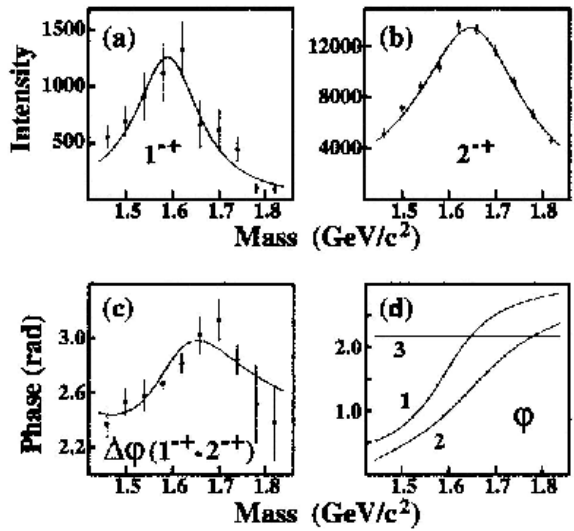}
 \caption{\it
The exotic $\pi_1(1600)$ observed 
by the VES and E852 Collaborations, here shown in E852
$\pi^- P \to (\rho\pi)^- P$ data\cite{E852_rhopi}.
    \label{fig3} }
\end{figure}

We now have strong evidence for a true 
$J^{PC}=1^{-+}$ 
exotic
at 1.6~GeV in $\rho\pi$ at BNL\cite{E852_rhopi} and VES\cite{VESmodes}
(see Fig.3 for the $\rho\pi$ mode),
and $\eta'\pi$ and $b_1\pi$ at VES.\cite{VESmodes,VESpi2} 
In addition a rather lighter state
at 1.4~GeV in $\eta\pi$ has been reported by BNL and Crystal 
Barrel.\cite{E852_etapi,CB_etapi} Thus, 
experimental hadron spectroscopy may 
finally have 
found the hybrid mesons anticipated by theorists for about 25 years.
Of course there is an unresolved concern that these 
experimental masses are somewhat lighter than the theoretical expectation
of
$\approx 1.9$-$2.0$~GeV.
There are also nonexotic hybrid candidates such as the 
$\pi(1800)$;\cite{VES_pi1800}
a recent and reasonably complete review of light
meson spectroscopy which discusses hybrid candidates in more
detail was recently completed by Godfrey and Napolitano.\cite{GodNap}.

Hybrid strong decays are in a confused state. The flux-tube model predicts
that the dominant modes should be S+P two-body combinations such as
$\pi f_1$
and
$\pi b_1$.\cite{IKP} 
The reported observations of hybrids however have
for the most part been in the more familiar S+S modes such as
$\pi\eta$, 
$\pi\eta'$ and
$\pi\rho$, although there is some evidence for
$\pi b_1$\cite{VESmodes}
and
$\pi f_1$.\cite{E818_pif1}
VES has reported relative branching fractions for the $\pi_1(1600)$
exotic hybrid
candidate that 
actually suggest comparable branching fractions to S+S and
S+P modes.\cite{VESmodes} 
Clearly the modelling of strong decays of hybrids is at an early
stage, and the experimental determination of relative $\pi_1$
hybrid branching fractions 
will be a very useful contribution 
(assuming that these states persist with 
improved statistics!).
 
Since $q\bar q g$ hybrids span flavor nonets, there should be many more
hybrids near 1.5~GeV 
if 
the reports of $\pi_1$ exotic hybrids near this mass are correct.
Specific models of hybrids such as the flux-tube and bag models
find that the 
majority of light hybrids have nonexotic 
$J^{PC}$. 
In the flux tube model the lightest hybrid multiplet contains 
five nonexotic quantum numbers,
\begin{equation}
J^{PC}({\rm lightest\  flux-tube\ hybrid\ nonexotics}) =
0^{-+}, 1^{--}, 1^{++}, 1^{+-}, 2^{-+}
\label{eq3} 
\end{equation}
whereas in the bag model the lightest hybrid multiplet contains
just three nonexotics,
\begin{equation}
J^{PC}({\rm lightest\  bag-model\ hybrid\ nonexotics})  =
0^{-+}, 1^{--}, 2^{-+}
\ .
\label{eq4} 
\end{equation}

Note that both models include a 
$1^{--}$ 
flavor nonet in the set of
lowest-lying hybrid mesons. Thus the 
$1^{--}$ 
sector should show evidence of overpopulation relative to the naive
quark model, which can be tested at DAFNE. We shall return to this topic 
in the next section.

\subsection{Multiquarks and Molecules}

In the 1970s it was thought that the existence of many basis states in the
$q^2\bar q^2$ sector implied a very rich spectrum of multiquark resonances.
Calculations in specific models such as the MIT bag model and color-truncated
potential models appeared to support this picture. However it was 
subsequently realized that the overlap of these multiquark basis states 
with the continuum of two color-singlet $(q\bar q)(q\bar q)$ 
mesons implied that the multiquark systems need not appear as resonances;
they might instead simply be components of nonresonant two-meson continua.

An exception to this absence of four-quark resonances
can occur if the multiquark system lies well below all
two-body decay thresholds, or if there is a strongly suppressed coupling 
to the open decay channels; 
in these cases we might still expect to identify a bag-model ``cluster" 
multiquark resonance. 

Nature appears to favor a different type of 
multiquark system, in which 
largely
unmodified color-singlet $q\bar q$ or $qqq$ hadrons are weakly bound by the
residual nuclear forces between color singlets. 
Examples of such quasinuclear multiquark systems abound; 
the table of nuclear
species gives far more examples than we have of individual hadrons,
and hypernuclei extend these systems into strangeness. In the mesonic
sector, however, just two possible examples are widely cited, the
scalar mesons $f_0(980)$ and $a_0(980)$.

These scalars are candidates for weakly bound
$K\bar K$ nuclei, ``molecules",\cite{Wei90} due to their
masses and quantum numbers (which are those of an S-wave $K\bar K$ pair),
and also because their hadronic couplings appear bizarre for $n\bar n$ states,
which should be very broad and for $I=0$ should couple strongly to $\pi\pi$.
Another problem with a conventional assignment is the two-photon widths
of these states, which are much smaller than expected for $q\bar q$
but are rather similar to predictions for $K\bar K$ bound states\cite{TBgams}
or 
$ns\bar n\bar s$ 
four-quark 
clusters.\cite{Achgams}
An interesting test of the nature of these states was proposed by
Close, Isgur and Kumano;\cite{CIK} the theoretical
radiative branching fractions 
from the
$\phi$ depend rather strongly on the 
quark model assignments, and 
for $q\bar q$ versus $K\bar K$ states are

\begin{equation}
B(\phi\to\gamma f_0(980), \gamma a_0(980)) =
\left\{
\begin{array}
{r@{\quad:\quad}l}
4 \cdot 10^{-5}   &  K\bar K\ {\rm{ (both\ states)}}    \\
\simeq 1 \cdot   10^{-5}   &  f_0(980) = s\bar s   \\
\leq     10^{-6}   &  f_0(980), a_0(980) = n\bar n  \ . 
\end{array}
\right.
\label{eq5} 
\end{equation}
Close {\it et al.} 
note that 
the ratio
$\phi\to\gamma a_0(980) /
\gamma f_0(980)$
is also of interest, since it can distinguish between different 
multiquark spatial wavefunctions. For a $K\bar K$ molecule
this ratio is 1, whereas for an $(ns)(\bar n\bar s)$ system
it is 9.\cite{CIK}

At this meeting we have heard that the new 
experimental results from VEPP\cite{Milstein} 
are not far from
the 
Close {\it et al.} 
predictions 
for a $K\bar K$ molecule. (The VEPP experimental branching fractions
$B(\phi\to\gamma f_0(980), \gamma a_0(980))$ 
are somewhat larger than
$4 \cdot 10^{-5}$, but are roughly consistent with 
Close {\it et al.} 
given the
current errors.) Earlier experimental indications of 
much larger branching
fractions to the 980~MeV states
were biased by large nonresonant contributions well below
980~MeV, which had not clearly been identified.

Presumably there are many meson-meson bound states, since 
many other meson pairs 
experience attractive residual nuclear interactions. 
Unlike glueballs and hybrids,
the spectrum of molecular states beyond 
$K\bar K$ and the nuclei and hypernuclei 
has received little
theoretical attention. 
There are quark model and meson-exchange model
predictions that 
some vector meson pairs may bind,\cite{vecmolec,deuson} 
but to date there has been little systematic investigation of
the expected spectrum. As our understanding of residual hadronic forces
improves, we can expect this to be one of the interesting areas of development
in hadron spectroscopy in the coming years.

\begin{figure}[t]
\vspace{9.0cm}
\includegraphics{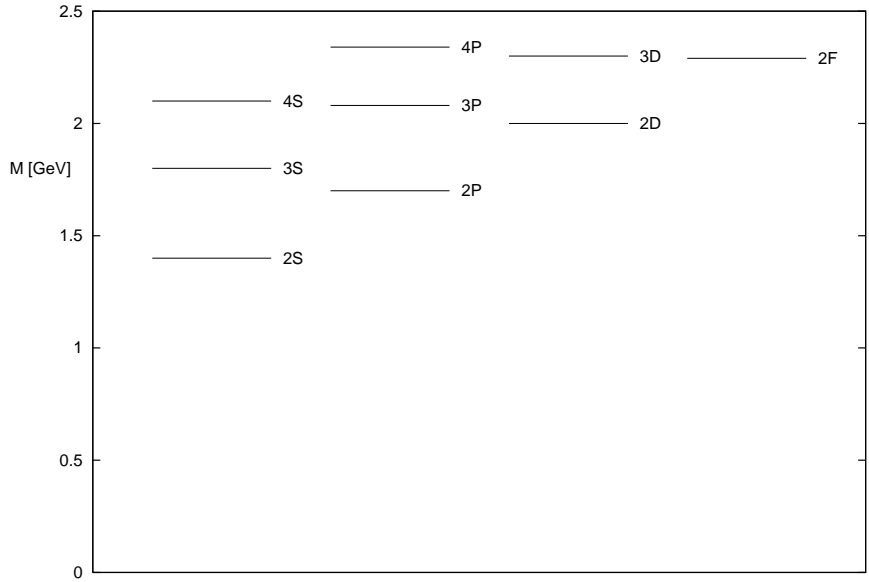}
\caption{\it
Excited $n\bar n$ multiplets suggested by recent data
(see Table~1). 
\label{fig4}   }
\end{figure}

\begin{table}
\centering
\caption{ \it 
Suggested 
excited 
$n\bar n$ 
multiplets. 
}
\vskip 0.1 in
\begin{tabular}{|l|c|l|} \hline
 nL         &  $M$(GeV) & representative WHS99 candidates \\
\hline
\hline
 2S         & $  1.4  $  &  $\rho(1450),\pi(1300)$    \\   
 3S         & $  1.8  $  &  $\pi(1740)$    \\   
 4S         & $  2.1  $  &  $\rho(2150)$    \\   
\hline
 2P         & $  1.7  $  &  $f_2(1650),a_2(1700),a_1(1700)$    \\   
 3P         & $  2.08 $  &  $f_0(2095),a_1(2100),a_0(2050)$   \\   
 4P         & $  2.34 $  &  $f_0(2335),a_1(2340)$   \\   
\hline
 2D         & $  2.0  $  &  $\omega_3(1950),\eta_2(2040)$ \\   
 3D         & $  2.3  $  &  $\rho_3(2300),\omega_3(2215),\eta_2(2300)$ \\   
\hline
 2F         & $  2.29 $  &  $f_4(2290),f_3(2280),a_4(2280),a_3(2310)$ \\   
\hline
\end{tabular}
\label{table1}
\end{table}

\subsection{Conventional $q\bar q$ Mesons}

As a background to these various hadronic exotica we have 
a spectrum of conventional $q\bar q$ states, which must be identified
if we are to isolate non-$q\bar q$ states. Since 
many of the light 
non-$q\bar q$ states 
predicted by theorists 
have masses and quantum numbers
that allow confusion with excited $q\bar q$ states, it is 
important to establish the light $q\bar q$ 
spectrum below 2.5~GeV as completely as possible.

Identification of the $q\bar q$
and non-$q\bar q$ states in the spectrum will require that we clarify meson
spectroscopy to a mass of at least 2.5~GeV, so that the pattern of glueballs,
hybrids and multiquarks can be established through the identification
of sufficient examples of each type of state.

There has been impressive experimental progress in the identification
of the (presumably $q\bar q$) light meson spectrum in recent years. In Fig.4 we
show the masses of the relevant
radially- and orbitally-excited multiplets for which
candidate states were reported at the WHS99 hadron spectroscopy 
meeting in Frascati earlier this year, from a review by Barnes.\cite{tb_WHS99}
It appears that almost all the $q\bar q$ multiplets
expected below 2.5~GeV 
have now been identified.\cite{extrarho}
These multiplet masses and some representative
candidates reported at the WHS99 meeting are given in Table~1. 

Surprisingly, these orbital+radial multiplets 
lie at rather lower masses than predicted by
Godfrey and Isgur;\cite{God85} compare the predicted and
observed 2P and 2D multiplet masses:

\begin{equation}
M(2P)|_{\rm GI} \approx 1.80 {\ \rm GeV},
\label{eq6} 
\end{equation}
\begin{equation}
\ 
M(2P)|_{\rm expt.} \approx 1.7 {\ \rm GeV}.
\label{eq7} 
\end{equation}
\begin{equation}
M(2D)|_{\rm GI} \approx 2.14 {\ \rm GeV},
\label{eq8} 
\end{equation}
\begin{equation}
M(2D)|_{\rm expt.} \approx 2.0 {\ \rm GeV}.
\label{eq9} 
\end{equation}
Evidently, experiment is 
finding the 2P and 2D multiplets
about 0.1-0.2~GeV lower in mass than predicted by the
Godfrey-Isgur model. If this discrepancy 
is confirmed it will be important to
determine whether this requires some important modification
of the model.

Thus far it has been possible to identify these $q\bar q$ multiplets largely
by the systematics of masses. This is possible because 
multiplet splittings
decrease rapidly with increasing $L$, so we are fortunate to find 
the 
members of a given
higher-$L$ multiplet
at very similar masses. 
In principle one might also distinguish between $q\bar q$ states
and non-$q\bar q$ exotica such as glueballs and hybrids through
their strong decay branching fractions and amplitudes. Detailed predictions
are now available for these branching fractions for all $n\bar n$ states
expected 
up to 2.1~GeV,\cite{bcps} and for a few specific cases at higher
mass.\cite{Blu} If our decay models are accurate, these higher quarkonia
often have very characteristic branching fractions, which should 
be quite distinct from glueball or hybrid decays.
Unfortunately, the 
$^3$P$_0$ decay model and the closely related flux-tube decay model
have not been tested carefully, except in a few cases such as
$b_1\to \omega\pi$
and
$a_1\to \rho\pi$. 
(These transitions have both S and D amplitudes, and their D/S ratios are
sensitive tests of the decay models and
are in good agreement with experiment.)
A new and very important test of the decay models was recently reported by 
VES.\cite{VESpi2} In both the $^3$P$_0$ and flux-tube decay models, transitions
of the type $(S_{q\bar q}=0)\to (S_{q\bar q}=0) + (S_{q\bar q}=0)$ are 
forbidden, due to the spin-1 character of the decay model pair creation
operator. This implies for example that $\pi_2(1670)\to b_1\pi$ should
vanish, although it is nominally an allowed D-wave strong decay.
VES finds a rather tight upper limit on this
transition,
\begin{equation}
B(\pi_2(1670)\to b_1\pi) < 0.19\% \ (2\sigma \ c.l.)\ .
\label{eq10} 
\end{equation}
This null result is very reassuring, but does not uniquely
confirm a $^3$P$_0$-type decay model; the same theoretical zero
follows for example from OGE pair production.\cite{abs}
A second test due to VES which also involves the $\pi_2(1670)$ 
does {\it not} agree with the expectations of the decay models: 
$B(\pi_2(1670)\to \omega\rho)$ should be about $16\%$,\cite{bcps} and
the spin-1 decay operator implies that the $\omega\rho$ final state
should have spin-1, with the $^3$P$_2$ $\omega\rho$ amplitude dominant.
VES instead finds
\begin{equation}
B(\pi_2(1670)\to \omega\rho \ (S=2) ) = 1.9(0.4)(1.0)\% 
\label{eq11}
\end{equation} 
and
\begin{equation}
B(\pi_2(1670)\to \omega\rho \ (S=1) ) = 0.9(0.2)(0.3)\%  \ ,
\label{eq12}
\end{equation} 
which suggests that strong decays in this sector may not agree
with the decay models. 

Until such time as we can test the predictions of the decay models
against a wide range of 
accurately determined experimental decay amplitudes and branching
fractions, it will remain unclear whether the predictions are indeed reliable,
or accidentally happen to work well for a few special cases.
For this reason it would be extremely useful to determine the relative
branching fractions of all two-body modes of higher-mass states
such as the excited vectors $\rho(1465)$ and $\rho(1700)$. The current
situation, with most modes unmeasured or reported only as ``seen"
(Tables 2-4) 
does not allow one to make progress in the very important subject of
strong decays.
An accurate determination of excited vector decay amplitudes
would be an extremely useful DAFNE contribution, as we shall now discuss.

\section{Exotica and excited vector mesons at DAFNE}

The ``vector sector" affords a very interesting subject for future 
investigation at DAFNE. This topic was studied 
at Frascati in the past at ADONE,\cite{ADONE} 
{\it albeit} with much lower luminosity.
$e^+e^-$ annihilation is of course the
ideal technique for making these states,
since single photons make 
$1^{--}$ uniquely.
At the time this appeared to be a rather straightforward
problem in hadron spectroscopy,
since
the quark model predictions of 
excited 
$J^{PC}=1^{--}$ 
vector mesons 
with 
radial 
and orbital 
excitations
(with both 2$^3$S$_1$ and $^3$D$_1$ expected at about $1{1\over 2}$~GeV)
were uncontroversial. 
Unfortunately it was found that the excited vectors 
were rather broad, overlapping states, so the 
radial and orbital 
excitations
could not be clearly separated. 
This subject has been reviewed by Donnachie,\cite{Donnachie,DK} who 
discusses it in more detail in these proceedings.

The subject advanced somewhat with studies of 
the $\rho$-sector in
both $2\pi$ and $4\pi$ modes,
which lead the PDG in 1988 to distinguish
two states,
the $\rho(1450)$ and the $\rho(1700)$. 
(The current status of light vector spectroscopy according to the PDG is
shown in Fig.5.) 
These are usually indentified with the 2S (radial) and D (orbital)
excitations respectively, since the masses correspond approximately
to quark potential model expectations. (There are problems with this
simple 
assignment, such as the surprisingly large $e^+e^-$ coupling of the
nominally $L=2$ $\rho(1700)$, which has a vanishing wavefunction at contact.) 

\begin{figure}[t]
\vspace{9.0cm}
\includegraphics{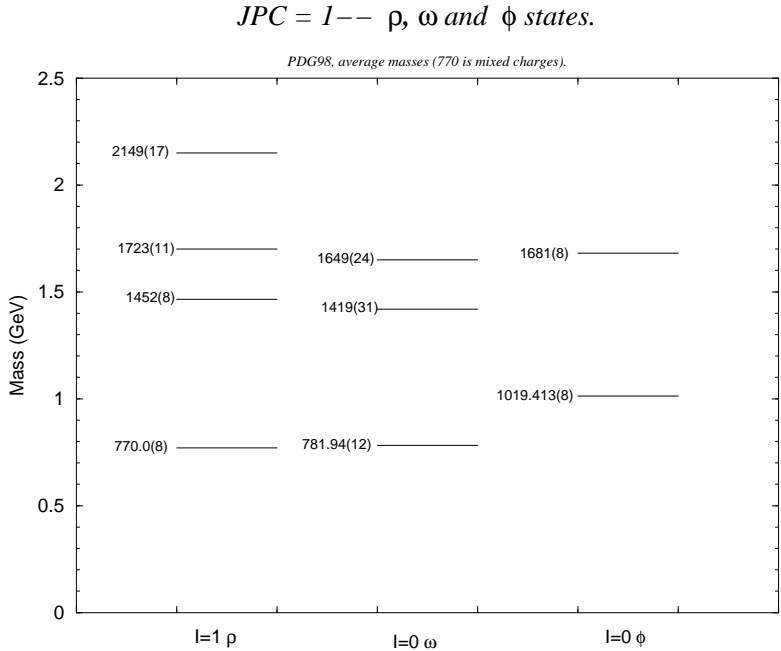}
\caption{\it Excited vector meson spectroscopy according to the 1998 PDG. 
\label{fig5}   }
\end{figure}

\begin{table}
\centering
\caption{\it Theoretical\cite{bcps,cp} and 
experimental\cite{PDG98} 
two-body partial widths and total widths 
(MeV) and branching fractions 
of excited $\rho$ states. Theoretical predictions
are for a
$\rho_{2S}(1465), \rho_D(1700)$ and a $\rho_H(M_H)$, with
$M_H=1.5$~GeV and $2.0$~GeV. For the $\rho_H(2000)$, $a_2\pi$ and
$K_1K$ modes (not shown) are also important.
}
\vskip 0.1 in
\begin{tabular}{|l|c|c|c|c|c|c|c|c|c|} 
\hline
$\Gamma_i$:
& $\pi\pi$  
& $\omega\pi$  
&  $\rho\eta$  
& $\rho\rho$  
&  $KK$  
&  $K^*K$  
&  $h_1\pi$  
&  $a_1\pi$  
&  $\Gamma_{tot}$  
\\
\hline
\hline
$\rho_{2S}$
&  74
& 122
&  25
&   -
&  35
&  19
&   1
&   3
& 279
\\
\hline
$\rho_D$
&  48
&  35
&  16
&  14
&  36
&  26
& 124
& 134
& 435
\\
\hline
$\rho_H(1500)$
&   0
&   5
&   1
&   0
&   0
&   0
&   0
& 140
& $\approx 150$
\\
\hline
$\rho_H(2000)$
&   0
&   8
&   7
&   0
&   0
&   4
&   0
& 170
& $\approx 340$
\\
\hline
\hline
B$_{expt.}$:
&  
& 
& 
& 
& 
& 
& 
& 
& 
\\
\hline
$\rho(1465)$
&  seen
& $< 2\%  $
& $< 4\%  $
&   -
& $< 0.16\%  $
&   -
&   -
&   -
& 310(60)
\\
\hline
$\rho(1700)$
&  seen
&  seen
&  seen
&  -
&  seen
&  seen
&  ${\rho\pi\pi \atop  dom.}$
&  ${\rho\pi\pi \atop  dom.}$
&  240(60) 
\\
\hline
\end{tabular}
\label{table2}
\end{table}

\begin{table}
\centering
\caption{\it As Table 2 but
for excited $\omega$ states. Theoretical predictions
are for an
$\omega_{2S}(1419), \omega_D(1649)$ and an $\omega_H(M_H)$,
with
$M_H=1.5$~GeV and $2.0$~GeV.}
\vskip 0.1 in
\begin{tabular}{|l|c|c|c|c|c|c|c|c|c|} 
\hline
$\Gamma_i$:
& $\rho\pi$  
& $\omega\eta$  
& $\omega\eta'$  
&  $b_1\pi$  
&  $KK$  
&  $K^*K$  
&  $K^<_1K$  
&  $K^>_1K$  
&  $\Gamma_{tot}$  
\\
\hline
\hline
$\omega_{2S}$
& 328 
&  12
&   -
&   1
&  31
&   5
&   -
&   -
& 378
\\
\hline
$\omega_D$
& 101
&  13
&   -
& 371
&  35
&  21
&   -
&   -
& 542
\\
\hline
$\omega_H(1500)$
&  20
&   1
&   -
&   0
&   0
&   -
&   -
&   -
& $\approx 20$ 
\\
\hline
$\omega_H(2000)$
&  40
&  20 
&  30 
&   0
&   0
&  30 
&  40 
&  60
& $\approx 220$ 
\\
\hline
\hline
B$_{expt.}$:
&  
& 
& 
& 
& 
& 
& 
& 
& 
\\
\hline
$\omega(1419)$
&  dom.
& 
& 
&   
& 
&   
&   
&   
& 174(59)
\\
\hline
$\omega(1649)$
&  seen
&  
&  
&  ${\omega\pi\pi\atop seen}$
&  
&  
& 
&  
&  220(35) 
\\
\hline
\end{tabular}
\label{table3}
\end{table}

\begin{table}
\centering
\caption{\it As Table 2 but
for excited $\phi$ states. Theoretical predictions
are for a
$\phi_{2S}(1680), \phi_D(1850)$ and a $\phi_H(2150)$.
}
\vskip 0.1 in
\begin{tabular}{|l|c|c|c|c|c|c|c|c|c|} 
\hline
$\Gamma_i$:
&  $KK$  
&  $K^*K$  
&  $K^*K^*$  
&  $K^<_1K$  
&  $K^>_1K$  
&  $K_2K$  
& $\phi\eta$  
& $\phi\eta'$  
&  $\Gamma_{tot}$  
\\
\hline
\hline
$\phi_{2S}$
& 89
&  245
&   -
&   -
&   -
&   -
&  44
&   -
& 378
\\
\hline
$\phi_D$
&  65
&  75
&   5
&   465
&   -
&   -
&   29
&   -
& 638
\\
\hline
$\phi_H(2150)$
&   0
&   15
&   0
&   60
&  125
&   20
&    8
&    2
& $\approx 230$ 
\\
\hline
\hline
B$_{expt.}$:
&  
&   
& 
& 
& 
& 
& 
& 
& 
\\
\hline
$\phi(1680)$
&  seen  
&  dom.
& 
& 
& 
&   
&   
&   
& 150(50)
\\
\hline
\end{tabular}
\label{table4}
\end{table}

There are analogous
states reported in the isosinglet sector, the
$\omega(1420)$ 
and 
$\omega(1600)$. 
(The situation may be more complicated.
See in particular the recent results 
on
$e^+e^-\to \pi^+\pi^-\pi^o$ 
from VEPP,\cite{VEPP_omega} 
which show a very low mass peak at about 1220~MeV.)
In the $\phi$ sector we have evidence for only a 
single excitation, the $\phi(1680)$. 

Interest in the excited vectors has increased with the realization
that the lightest hybrid meson multiplet 
includes a $1^{--}$
(in both the flux-tube and bag models),
and that these hybrid vectors are predicted to be rather narrow.
Indeed, in the hybrid meson decay calculations of Close and Page 
(using the Isgur-Kokoski-Paton flux-tube model) the narrowest hybrid
found was the $\omega$-flavor $1^{--}$. 
(See Table 3 for the predicted partial widths of this vector hybrid.) 
The Close-Page calculations assumed
a mass of 2.0~GeV for the 
$\rho_H$ and $\omega_H$
hybrid vectors, but given the reports of the
$\pi_1(1400)$ and $\pi_1(1600)$ hybrid candidates,
one should also
consider the possibility that the lowest hybrid multiplet lies at about
1.5~GeV. This would give us a third $1^{--}$ level roughly degenerate with
the quark model 2S and D levels, and such light vector hybrids could be
very narrow (see Table~3); a hypothetical $\omega_H(1500)$ is predicted to have 
a total width of only
about 20~MeV!
If the $1^{--}$ hybrid states are not found at this low
mass, one might question the reports of $\pi_1$ $1^{-+}$ exotics near 1.5~GeV.

The topic of vector meson spectroscopy in this mass region
was recently reviewed by Donnachie and 
Kalashnikova,\cite{DK} who concluded that additional vectors
beyond the expected $q\bar q$ states
are indeed required 
to fit the data
in both $I=0$ and $I=1$ channels.
In $I=1$ in particular, the 
weakness of
$e^+e^-\to \pi^+\pi^-\pi^o\pi^o$ relative to
$e^+e^-\to \pi^+\pi^-\pi^+\pi^-$ cannot be
explained by the expected $\rho(1700)$
decays alone. 

In principle one should be able to separate the 2S, D and H (hybrid) states
by studies of their relative decay branching fractions. In Tables 2-4 we 
show theoretical predictions for the different types of states, compared
to 1998 PDG results for experimental branching fractions. 
The relative strength of the broad $4\pi$ modes
$h_1\pi$ and $a_1\pi$ is quite sensitive to the type of parent resonance,
and could serve as a useful discriminator if the decay models are accurate.
The theoretical expectation is that the D state should populate 
both modes, H should only populate $a_1\pi$, and the 2S state 
does not couple significantly to either of these modes.
How well do these theoretical predictions agree with experiment?

The experimental branching fractions of excited vectors, 
as reported in the 1998 PDG, are also shown
in Tables 2-4. It
is clear at a 
glance that experiment is in a woeful state. Almost nothing is known about
the decays of excited $\omega$ and $\phi$ states. 
(Note that excited $\phi$ vectors can be isolated by studying the
$s\bar s$-filter mode 
$\phi\eta$, which was apparently not attempted previously.)
In the $\rho$ sector,
there are promising indications that the $\rho(1700)$ may be observed in
many of the expected channels, but there is almost no quantitative 
information about relative branching fractions which we require for
tests of the decay models. In contrast, there are strong limits claimed
for $\rho(1465)$ branching
fractions, which 
appear to be very different from expectations for a simple
2S radial excitation. Note especially the tight limit 
$B(\rho(1465)\to\omega\pi) < 2\% $. Taken literally, this result 
is very interesting in that it argues strongly
against a 2S assignment for the $\rho(1465)$. (Compare the 
$\rho_{2S}$ 
and
$\rho(1465)$ entries in Table~2.)
Unfortunately it is difficult to reconcile this number
with the
reported {\it dominance} of $\omega(1419)\to \rho\pi$ (Table~3), 
since that decay differs from $\rho(1465)\to\omega\pi$ only
by a flavor factor  
of 3  
(favoring $\omega(1419)\to \rho\pi$) and minor changes in phase space.

Recently the Crystal Barrel Collaboration attempted to separate the
contributions of the $\rho(1465)$ and $\rho(1700)$ to the various
$4\pi$ final states. Initially the results appeared consistent with
the usual quark model assignments 2S and D,\cite{Thoma97} 
but the most recent 
work\cite{Pick99} has found that essentially all broad $4\pi$ modes
($a_1\pi, h_1\pi, \pi(1300)\pi, \rho\rho$ and $\rho\sigma$) are important
in the decays of both the $\rho(1465)$ and the 
$\rho(1700)$! Unfortunately the statistical errors of this
many-parameter fit are rather large, so each mode typically has a 
fitted branching fraction 
about $2\sigma$ from zero. The excited vectors 
would evidently benefit from a study at an $e^+e^-$ facility such as DAFNE, 
where the complication 
of competing amplitudes in many other $J^{PC}$ channels is not present.

In view of the poorly constrained and perhaps inconsistent 
branching fractions evident in the PDG,
the most reasonable approach in future would probably be to 
study as many of the quasi-two-body decay modes in Tables 2-4 
as possible, determine
numerical values for the relative branching fractions, and carry out
a global fit of each flavor sector 
with an assumed two versus three parent resonances in each flavor.

\section{Two-photon couplings}

In the opinion of at least two LEAR experimentalists,\cite{KleTho} 
using $\gamma\gamma$ collisions
to clarify the scalar sector is the most 
interesting contribution DAFNE
could make to spectroscopy.

Two-photon couplings of resonances can be inferred by measurement of
the cross section
\begin{equation}
\sigma(e^+e^- \to e^+e^- R)
\label{eq13} 
\end{equation}
which is proportional to the two-photon width $\Gamma{\gamma\gamma}$
of the resonance $R$,
as discussed in Sec.36.3 of the 1998 PDG.\cite{PDG98}
Two-photon widths of $C=(+)$ resonances have been measured
at several $e^+e^-$ facilities in the past, most recently
at LEP.\cite{Aleph,L3} These are especially interesting quantities 
because they show considerable variation between $q\bar q$ and non-$q\bar q$
states, and if determined with sufficient accuracy they could be used
for example to solve the problem of the assignments of the 
various light scalars. This subject attracted considerable interest and
effort previously, but as $e^+e^- \to e^+e^- R$ is an $O(\alpha^4)$ process
and the cross section falls rapidly with $M_R$, it was not possible
to obtain adequate statistics for a definitive analysis.

The two-photon partial widths of $q\bar q$ states within a flavor multiplet
in the SU(3) limit are in the ratio
\begin{equation} 
\Gamma{\gamma\gamma}\ \ \ \ f\ :\ a\ :\ f'\ =\  25\ :\ 9\ :\ 2\ \ ,
\label{eq14} 
\end{equation} 
so if a candidate $q\bar q$ state such as the $2^{++}$ $f_2(1270)$ 
is reported, one should also observe its flavor partners at about this
relative strength. For example, the $\Gamma{\gamma\gamma}$ widths of the
$2^{++}$ multiplet are 
\begin{equation}
\Gamma{\gamma\gamma}(2^{++})\ \ \  f_2(1270) : a_2(1310) : f'(1525)\ =\ 
2.8(4) {\rm keV} :\ 1.00(6) {\rm keV} :\ \approx 0.1 {\rm keV} \  .
\label{eq15} 
\end{equation}     
(Moderate suppression of the $s\bar s$ coupling is expected
theoretically due to the heavier
strange quark mass.)

Scalars are predicted to have very characteristic two-photon couplings.
The largest $\Gamma{\gamma\gamma}$ width expected for any $q\bar q$ meson
is for the $^3$P$_0$ $f_0$ scalar; in the nonrelativistic quark model
it has a $\Gamma{\gamma\gamma}$ width 15/4 times that of the $f_2$,
and with relativistic corrections\cite{f0gamsthy} the ratio is reduced
to $\approx 2$. Thus for a scalar $n\bar n$ partner of the $f_2(1270)$
we expect a two-photon width of about 5 keV. An $f_0(1250)$ 
scalar signal of about
this strength was observed by the Crystal Ball Collaboration 
in $\gamma\gamma\to\pi^o\pi^o$ at DESY,\cite{CBallf0}
and may be the long-sought and still obscure $n\bar n$ scalar. 
In contrast, a pure scalar glueball should have a much smaller
two-photon width, since it has no direct coupling to photons. 
The recent ALEPH results on $\gamma\gamma$ couplings of resonances
appear to support the $f_0(1500)$  as a glueball candidate, since their
upper limit\cite{Aleph}
\begin{equation}
\Gamma{\gamma\gamma}(f_0(1500))\ < \ 0.17\ {\rm keV}\ (95\% \ c.l.)
\label{eq16} 
\end{equation}
is far below the ca.~5~keV expected for an $n\bar n$ scalar.

The various $n\bar n \leftrightarrow G \leftrightarrow s\bar s$
mixing models in contrast would predict $\Gamma{\gamma\gamma}$ widths
roughly proportional to each state's $n\bar n$ amplitude squared, and so
could be tested by the relative strength of each scalar resonance in 
$\gamma\gamma\to\pi^o\pi^o$. Finally, $K\bar K$ molecules\cite{TBgams}
and multiquark states\cite{Achgams} are predicted to have much smaller 
$\Gamma{\gamma\gamma}$ widths than the corresponding $n\bar n$ states,
which is in agreement with the sub-keV $\Gamma{\gamma\gamma}$ 
values reported
for the $f_0(980)$ and $a_0(980)$.

In contrast with the non-observation of the scalar glueball
candidate $f_0(1500)$ in $\gamma\gamma$, 
we now have clear evidence for 
the pseudoscalar $\eta(1440)$ in
$\gamma\gamma\to K_s K^{\pm} \pi^{\mp}$, reported by 
the L3 Collaboration.\cite{L3}
Once a glueball candidate
(this assignment is now implausible due to the high mass predicted 
for the pseudoscalar glueball by LGT), this state appears most
likely to be
a radially-excited $q\bar q$. Similarly there is a possible observation
of the scalar glueball candidate $f_0(1710)$ by L3 in 
$\gamma\gamma\to K_s K_s$, although this is preliminary. If the 
$f_0(1710)$ appears clearly 
in $\gamma\gamma$ at the rate expected for a radially-excited
$2^3$P$_0$ $n\bar n$ state, we may be able to eliminate it as a glueball
candidate in favor of the $f_0(1500)$. Clearly, accurate measurements 
of scalar $\Gamma{\gamma\gamma}$ couplings show great promise as a 
technique for solving the long standing problem of the nature of the
various $f_0$ scalar resonances.

\section{Acknowledgements}
It is a pleasure to acknowledge 
the kind invitation of
the organizers of the DAFNE meeting
to discuss the status of light meson
spectroscopy.
I would also 
like to thank 
my colleagues for 
discussions of various aspects of hadron physics in the 
preparation of this report,
in particular
N. Achasov, F.E. Close, A. Donnachie, U. Gastaldi, S. Godfrey, 
N. Isgur, Yu. Kalashnikova, E. Klempt,
S. Krewald, A.I.Milstein, 
C.J. Morningstar, P.R. Page, M.R. Pennington, B. Pick, 
J. Speth, E.S. Swanson, 
U. Thoma and N. T\"ornqvist.
Research at the Oak Ridge National Laboratory was 
supported by the U.S. Department of Energy under contract
DE-AC05-96OR22464 with Lockheed Martin Energy Research Corp.,
and additional support was provided by
the Deutsche Forschungsgemeinschaft under contract
Bo 56/153-1.

\end{document}